\documentclass[unnumsec,webpdf,contemporary,large]{oup-authoring-template}%
\graphicspath{{figures/}}

\theoremstyle{thmstyleone}%
\theoremstyle{thmstyletwo}%
\theoremstyle{thmstylethree}%

\usepackage{hyperref}

\usepackage{titlesec}
\definecolor{jnlclr}{cmyk}{0,.94,.91,0.38}
\titleformat*{\section}{\secsize\bfseries\color{jnlclr}}
\titleformat*{\subsection}{\subsecsize\color{jnlclr}}
\hypersetup{
    frenchlinks=true,
    colorlinks=true,
    linkcolor=jnlclr,
    citecolor=jnlclr,
    urlcolor=jnlclr
}

\begin{document}

\journaltitle{Preprint}
\DOI{DOI HERE}
\copyrightyear{2024}
\pubyear{2024}
\access{Advance Access Publication Date: Day Month Year}
\appnotes{Paper}

\firstpage{1}

%\subtitle{Subject Section}

\title[Streaming Large-Scale Electron Microscopy Data to a Supercomputing Facility]{Streaming Large-Scale Microscopy Data to a Supercomputing Facility}

\author[1,$\ast$]{Samuel S. Welborn\ORCID{0000-0002-7697-6347}} % chktex 8
\author[1]{Chris Harris\ORCID{0000-0002-1113-3728}} % chktex 8
\author[2]{Stephanie M. Ribet\ORCID{0000-0002-7117-066X}} % chktex 8
\author[2]{Georgios Varnavides\ORCID{0000-0001-8338-3323}} % chktex 8 
\author[2]{Colin Ophus\ORCID{0000-0003-2348-8558}} % chktex 8
\author[1,$\ast$]{Bjoern Enders\ORCID{0000-0002-6009-6281}} % chktex 8
\author[2,$\ast$]{Peter Ercius\ORCID{0000-0002-6762-9976}} % chktex 8 chktex 46

\authormark{Welborn et al.}

\address[1]{\orgdiv{NERSC}, \orgname{Lawrence Berkeley National Laboratory}, \orgaddress{\street{Berkeley}, \postcode{94720}, \state{CA}, \country{Country}}}
\address[2]{\orgdiv{NCEM, The Molecular Foundry}, \orgname{Lawrence Berkeley National Laboratory}, \orgaddress{\street{Berkeley}, \postcode{94720}, \state{CA}, \country{Country}}}

\corresp[$\ast$]{Corresponding authors. \href{email:swelborn@lbl.gov}{swelborn@lbl.gov}, \href{email:benders@lbl.gov}{benders@lbl.gov}, \href{email:percius@lbl.gov}{percius@lbl.gov}}  % chktex 46 chktex 12

%\editor{Associate Editor: Name}

%\abstract{
%\textbf{Motivation:} .\\
%\textbf{Results:} .\\
%\textbf{Availability:} .\\
%\textbf{Contact:} \href{name@email.com}{name@email.com}\\
%\textbf{Supplementary information:} Supplementary data are available at \textit{Journal Name}
%online.}

\abstract{Data management is a critical component of modern experimental workflows. As data generation rates increase, transferring data from acquisition servers to processing servers via conventional file-based methods is becoming increasingly impractical. The 4D Camera at the National Center for Electron Microscopy (NCEM) generates data at a nominal rate of 480 Gbit \(\mathrm{s}^{-1}\) (87,000 frames \(\mathrm{s}^{-1}\)), producing a 700 GB dataset in fifteen seconds. To address the challenges associated with storing and processing such quantities of data, we developed a streaming workflow that utilizes a high-speed network to connect the 4D Camera's data acquisition (DAQ) system to supercomputing nodes at the National Energy Research Scientific Computing Center (NERSC), bypassing intermediate file storage entirely. In this work, we demonstrate the effectiveness of our streaming pipeline in a production setting through an hour-long experiment that generated over 10 TB of raw data, yielding high-quality datasets suitable for advanced analyses. Additionally, we compare the efficacy of this streaming workflow against the conventional file-transfer workflow by conducting a post-mortem analysis on historical data from experiments performed by real users. Our findings show that the streaming workflow significantly improves data turnaround time, enables real-time decision-making, and minimizes the potential for human error by eliminating manual user interactions.}
\keywords{streaming, 4D-STEM, high-performance computing, real-time processing}

% \boxedtext{
% \begin{itemize}
% \item Key boxed text here.
% \item Key boxed text here.
% \item Key boxed text here.
% \end{itemize}}

\maketitle

\section{Introduction}

In the era of big data, the scientific community faces significant challenges in data management~\citep{rao2020deluge, Spurgeon2021-ym}. This is especially evident at experimental user and core facilities, where advancements in instrumentation, such as faster detectors and increased light source brightness, have led to an exponential increase in data generation rates.  The traditional methods of data storage and movement (e.g., personal flash drives) are becoming increasingly untenable.

In 2019, a new detector called the 4D Camera was installed on the TEAM 0.5 microscope at the National Center for Electron Microscopy (NCEM) facility of The Molecular Foundry at Lawrence Berkeley National Laboratory (LBNL)~\citep{ercius20234dcamera,ercius20204dcamera}. This detector produces data at a rate of 480 Gbit \(\mathrm{s}^{-1}\) (equivalent to 87,000 frames \(\mathrm{s}^{-1}\)), yielding datasets of up to 700 GB for a fifteen second acquisition. Other microscopy facilities are installing similar high frame rate detectors with the ability to routinely generate \(> 100\) GB datasets~\citep{chatterjee2021ultrafast, zambon2023high}. While these technological advancements provide new avenues for scientific exploration, they also pose significant challenges in data management, analysis, and acquisition. New opportunities for development include on-the-fly processing for quick feedback on an experimental approach and implementation of complex experimental pipelines, such as focal series or tomography~\citep{pelz2022solving,pelz2021smatrix} that leverage the capabilities of these advanced detectors. Given that microscope time is a limited and valuable resource, rapid data analysis that provides feedback on the quality of large data sets during a microscope session is crucial for improving throughput. 

To mitigate these challenges, a collaborative effort involving high performance computing (HPC) experts at the National Energy Research Scientific Computing Center (NERSC), electron microscopy experts at NCEM, and software development experts at Kitware, Inc.\ led to the utilization of NERSC for data reduction and the development of a web frontend called \textit{Distiller} to facilitate data management~\citep{distiller}.  HPC systems are typically accessed through command line interfaces, which are often unfamiliar to microscopists.  \textit{Distiller}, on the other hand, allows users to transfer and process data at NERSC through simple web-based interactions. This effort, which was part of a broader initiative at LBNL called The Superfacility Project, greatly improved the workflow for the 4D Camera~\citep{distiller, enders2020cross, welborn2024accelerating}. % chktex 12

Despite its utility, data analysis at NERSC was constrained by file-based input/output (I/O) steps that created bottlenecks at two stages: (1) writing data from random-access memory (RAM) to local disk storage at NCEM and (2) file transfer from NCEM to NERSC before computation. We note that file-based data movement is the common workflow across most detector systems. The dependence on file-based I/O operations slows down data processing, constrains the scope of feasible experiments, and relies on file systems (e.g., the NERSC global file system) possibly shared by multiple users. In our recent work, we showed that, by circumventing these file-based operations through streaming data from detector buffer memory directly to NERSC compute node memory, we improved throughput by five- to fourteen-fold~\citep{welborn2024accelerating}. In the present work, we showcase the advantages of a streaming workflow for microscopy experiments using the 4D Camera as a case study.

This manuscript is organized as follows. In the background section, we provide an overview of 4D scanning transmission electron microscopy (4D-STEM) and discuss difficulties in managing the substantial datasets generated by the 4D Camera. Then, we briefly outline the components of the streaming pipeline. Next, we describe enhancements to \textit{Distiller} that obviate the need for an in-depth understanding of HPC. Finally, we demonstrate the practical benefits of streaming through a comparative analysis of real user experiments employing both workflows. % chktex 13

\section{Background}\label{sec:background}

\subsection{Transmission Electron Microscopy and 4D-STEM}
Transmission electron microscopy (TEM) provides insights into the atomic and molecular structure of materials, making it a cornerstone characterization technique across scientific disciplines from materials science to biology. Scanning TEM (STEM) operates in a mode where an electron probe is focused onto the sample and rastered over a two-dimensional set of probe positions. Post-specimen detectors register electron events in diffraction space that can be mapped to specific probe positions. The versatility of STEM extends its utility beyond conventional imaging, facilitating advanced analytical methods such as spectroscopy, electron tomography, ptychography, and holography~\citep{yasin_harvey_chess_pierce_mcmorran_2016,yasin2018probing,stevens2018subsampled,ophus20194dstem, miao2016atomic, ercius2015electron, varnavides2023iterative, ribet2024uncovering,ben2021chain,ophus2023quantitative}.

Recent advancements in detector technology have ushered in a new era for STEM. Specifically, the introduction of direct electron detectors (DEDs) has dramatically accelerated data acquisition rates and opened new experimental possibilities ~\citep{levin2021, ercius20234dcamera}. 
DEDs can acquire data with a temporal resolution ranging from milliseconds to microseconds enabling a technique generally called 4D-STEM because two-dimensional (2D) diffraction patterns are acquired at a series of 2D probe positions~\citep{ophus20194dstem}. The resulting 4D dataset contains a wealth of both structural and compositional information about the sample. Analysis of the diffraction patterns can reveal the sample's overall crystal orientation, strain, and material phase, enabling a detailed mapping of these properties to provide a comprehensive characterization of the material~\citep{ophus20194dstem}. One of the applications of 4D-STEM is phase contrast imaging---while detectors record only the intensity of the exit wave after interaction with the sample, it is possible to reconstruct the phase, leading to dose-efficient characterization of weakly scattering signals. Phase retrieval STEM methods, such as differential phase contrast (DPC)~\citep{dekkers1974differential, waddell1979linear, shibata2012differential, cao2018theory}, which measures the change in the center of mass of diffraction patterns, and advanced algorithms such as ptychography, offer enhanced contrast and resolution~\citep{nellist1995resolution, varnavides2023iterative, enders2016ptypy}.

The size of 4D-STEM data introduce significant challenges in data management. An illustrative case is the 4D Camera, which can accumulate 2D diffraction patterns at a rate of 87,000 Hz (nominally 200 TB \(\mathrm{hr}^{-1}\)) highlighting the need for informed data treatment beyond current capabilities at most electron microscopy laboratories. Solving the challenges that come with storing and processing large datasets in a timely manner necessitates an examination of the pathway data takes within a data acquisition (DAQ) system and processing workflow.

\subsection{High Data Rate Acquisition and its Challenges}\label{sec:DAQ}

\begin{figure*}[!t]
  \centering
  \includegraphics[width=1\linewidth]{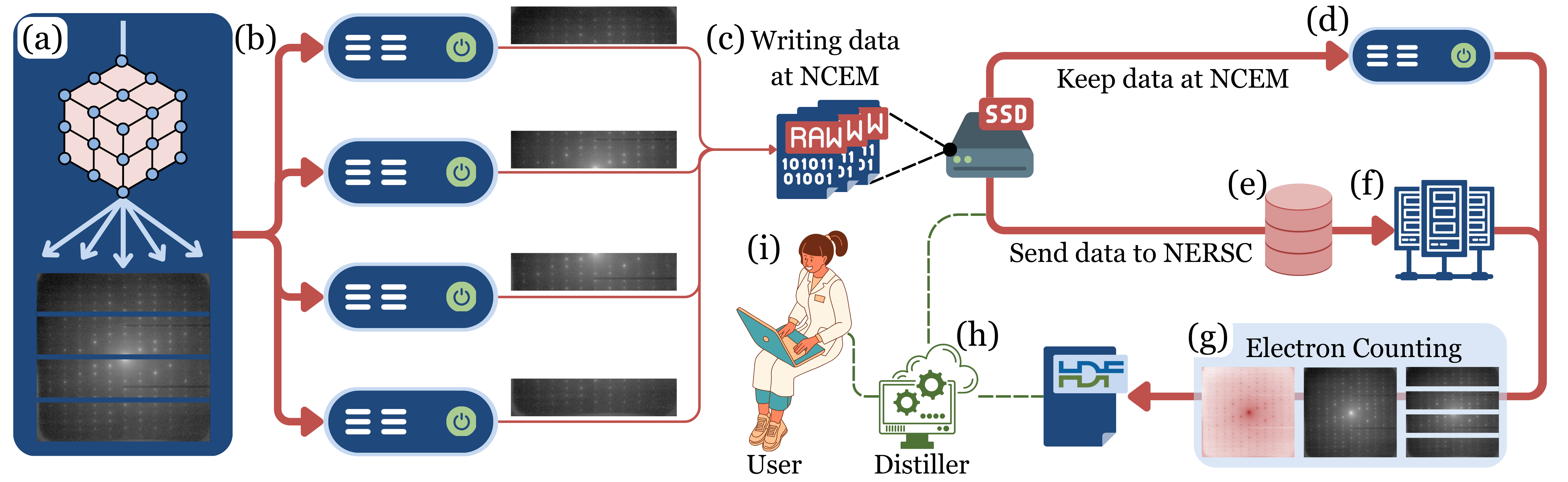}
  \caption{Schematic illustrating both the DAQ system and the initial mitigation strategies for managing large-scale 4D-STEM datasets generated at NCEM. A user begins an experiment using the TEAM 0.5 microscope software for the four-sector 4D Camera (a). The camera is connected to data receiving servers through FPGAs (b). Each server ingests all data into RAM and subsequently writes it to an eight TB flash storage system (c), which takes around 150 s for a 700 GB dataset. The data is either processed locally at NCEM on a single server with ten CPU cores (d), or transferred to NERSC's filesystems (e) and processed with more robust compute resources (f). Data processing is illustrated in panel (g), showing the assembly of disconnected sectors into coherent frames and subsequent electron counting of these frames. This processed data is saved in a single HDF5 file. The \textit{Distiller} web application (h) enables the user (i) to initiate file transfers to NERSC's file systems, perform electron counting, and launch analysis notebooks in NERSC's \textit{Jupyter} environment.}\label{fig:overview_prior_work} % chktex 13
\end{figure*}

The DAQ system for the 4D Camera at NCEM, developed in-house at LBNL, integrates both software and hardware elements to achieve such high data rates (Fig.\ \ref{fig:overview_prior_work}). The 4D Camera sensor (Fig.\ \ref{fig:overview_prior_work}{a}, bottom) is partitioned into four sectors, each of which is connected to a dedicated receiving server via twelve 10 Gbit \(\mathrm{s}^{-1}\) connections through field-programmable gate arrays (FPGAs). As the electron beam rasters across the sample (Fig.\ \ref{fig:overview_prior_work}{a}, top), a 576 \(\times\) 576 pixel frame is acquired at each scan position. Each 144 \(\times\) 576 pixel sector is processed by an FPGA and transmitted to its corresponding data receiving server (Fig.\ \ref{fig:overview_prior_work}{a-b}). Upon completion of a scan, the data are written as binary data files (Fig.\ \ref{fig:overview_prior_work}{c}) to flash storage. For a more comprehensive description of this DAQ system, the reader is referred to \cite{ercius20234dcamera} and \cite{welborn2024accelerating}. % chktex 21

With a data rate of 480 Gbits \(\mathrm{s}^{-1}\), a single fifteen second acquisition using the 4D camera generates approximately 700 GB of data~\citep{ercius20234dcamera}. The large data volume manifests three distinct but interrelated challenges: (1) limited local disk storage capacity, where the available eight TB of flash storage can only accommodate eleven full scans; (2) the computational burden associated with processing large datasets, which overwhelms local dedicated resources; and (3) the time-intensive nature of writing large files to disk, which blocks the system from further data acquisition. Collectively, these challenges substantially reduce user productivity and waste precious beam time. It is important to note that challenges in data management and computational limitations extend beyond NCEM to other Experimental and Observational Science (EOS) facilities, and these problems will intensify in the future~\citep{rao2020deluge, Spurgeon2021-ym}. A notable unscalable example is the Event Horizon Telescope data transfer protocol, which involved physically transporting hard disk drives from the telescope to a processing facility to produce the now-famous black hole image~\citep{galaxies}. 

\subsection{Initial Mitigation Strategies}\label{sec:mitigation_strategies}
The 4D Camera is designed to acquire CBED frames containing a small number of electrons, leading to a sparse data set. Thus, we can simultaneously mitigate the first challenge (storage capacity) and remove detector noise from our data through compression. The software package \textit{stempy}~\citep{stempy} efficiently transforms the raw data into a more manageable sparse format by finding and keeping only the locations of single electron hits, a process called ``electron counting'' in this manuscript~\citep{pelz2021eventrepresentation, ercius20234dcamera, Battaglia2009-cn}. This transformation (represented graphically in Fig.\ \ref{fig:overview_prior_work}{g}) results in an order of magnitude data size reduction (alleviating some storage pressure) and the reduction of detector noise. The raw detector data is typically deleted after it has been electron counted.

While \textit{stempy} significantly reduces storage requirements, it introduces the second challenge: computational demands for quickly processing large datasets. The processing time for this operation on local resources, represented in Fig.\ \ref{fig:overview_prior_work}{d}, is considerable. At NCEM, the computational resources are limited to ten CPU cores, which makes the electron counting of a 700 GB dataset a time-consuming task (10-12 minutes). During this time, the detector cannot be used because the same computational resources are shared for both data acquisition and processing. In contrast, each CPU node on NERSC's newest supercomputer, Perlmutter, is equipped with 128 CPU cores (Fig.\ \ref{fig:overview_prior_work}{f}), and multiple compute nodes can be allocated to parallelize the electron counting process. Upgrades to NERSC's computational infrastructure (which have occurred since the 4D Camera was installed) translate into immediate improvements in both the NCEM processing pipeline and for the broader NERSC user community, thereby optimizing resource utilization. Absent this integration, any dedicated compute nodes installed at NCEM require local maintenance and remain underutilized, particularly in periods between experiments. Moreover, by integrating NCEM's workflow with NERSC's infrastructure, NCEM users gain access to NERSC's rich computing and data ecosystem, which is particularly advantageous for processing their data during and after the experiment. This integration not only streamlines NCEM's operations but also provides a blueprint for the efficient deployment of compute resources beyond a single detector or EOS facility~\citep{enders2020cross, bard2022superfacility}. % chktex 8

% TODO: \ref needs to be fixed
Recognizing the advantages of centralized compute/storage resources for managing large datasets, the \textit{Distiller} (Fig.\ \ref{fig:overview_prior_work}{h}) application was developed to facilitate user interactions with the detector and NERSC. During data acquisition, \textit{Distiller} presents status and metadata using a user-friendly web-based frontend, allowing users (Fig.\ \ref{fig:overview_prior_work}{i}) to initiate data transfers to NERSC (Fig.\ \ref{fig:overview_prior_work}{e}). Then, the data is electron counted using \textit{stempy} on Perlmutter (Fig.\ \ref{fig:overview_prior_work}{f-g})~\citep{distiller,enders2020cross}. After counting, the end result is a single sparse HDF5 file ready for further analysis. NERSC can then provide access restrictions based on user credentials, compute for further analysis, and file transfer to other sites. We provide a screencast of this workflow in Supplemental Video 1~\citep{zenodo_videos}. It is also important to recognize that by collaborating with software development experts at Kitware and HPC specialists at NERSC, we avoided the technical debt often associated with \textit{ad hoc} scripts developed by microscopists, who do not typically have the bandwidth to develop seamlessly-integrated tools like \textit{Distiller}. % chktex 13

Despite these advancements, writing/reading large files to/from disk remains an unresolved bottleneck, leading to the third challenge that impedes the efficient transfer of high-volume data.

\subsection{I/O Bottlenecks in Data Transmission}\label{sec:bottlenecks}
Four critical I/O operations slow down the transmission of data from NCEM to NERSC: % chktex 13

\begin{enumerate}
    \item Writing the data to a local drive at NCEM. % chktex 13
    \item Reading the data from the local drive and transferring it to NERSC over a fiber network.
    \item Writing the data to NERSC's file systems.
    \item Reading the data into NERSC compute node memory for electron counting.
\end{enumerate}

These file I/O bottlenecks present a dual challenge: they slow down data transfer and analysis and also restrict the types of experiments that can be conducted. For instance, they preclude the possibility of running automated experiments over extended periods~\citep{Pattison2023-kt}, because human intervention is required to manage data transfer and counting once the local eight TB file system is full.

\section{Streaming Data from NCEM to NERSC}\label{sec:methods}

To overcome the I/O bottlenecks outlined above, we have developed a streaming service that facilitates the transmission of microscope data from NCEM servers to NERSC compute nodes without using file storage. The foundation of our solution is a socket-based network that facilitates RAM-to-RAM data transfer for real-time processing (Fig.~\ref{fig:benefits_streaming}). Sockets serve as integral components in networked systems, facilitating the exchange of data packets between interconnected devices; by using sockets, we are taking advantage of the progress made in commercial internet infrastructure to improve scientific computing. Our architecture utilizes Zero Message Queue (ZeroMQ), a network socket library, to establish communication between the key elements of our pipeline: the data receiving servers at NCEM, a centralized aggregator server at NCEM, and the compute nodes at NERSC. It is important to note that this section's content serves as a high-level synopsis of our approach. For a more comprehensive overview of the methods and system architecture, the reader is directed to our recent technical work~\citep{welborn2024accelerating}. % chktex 13

\subsection{Intercepting File Write at NCEM}
\begin{figure*}[!t]
    \centering
    \includegraphics[width=1\linewidth]{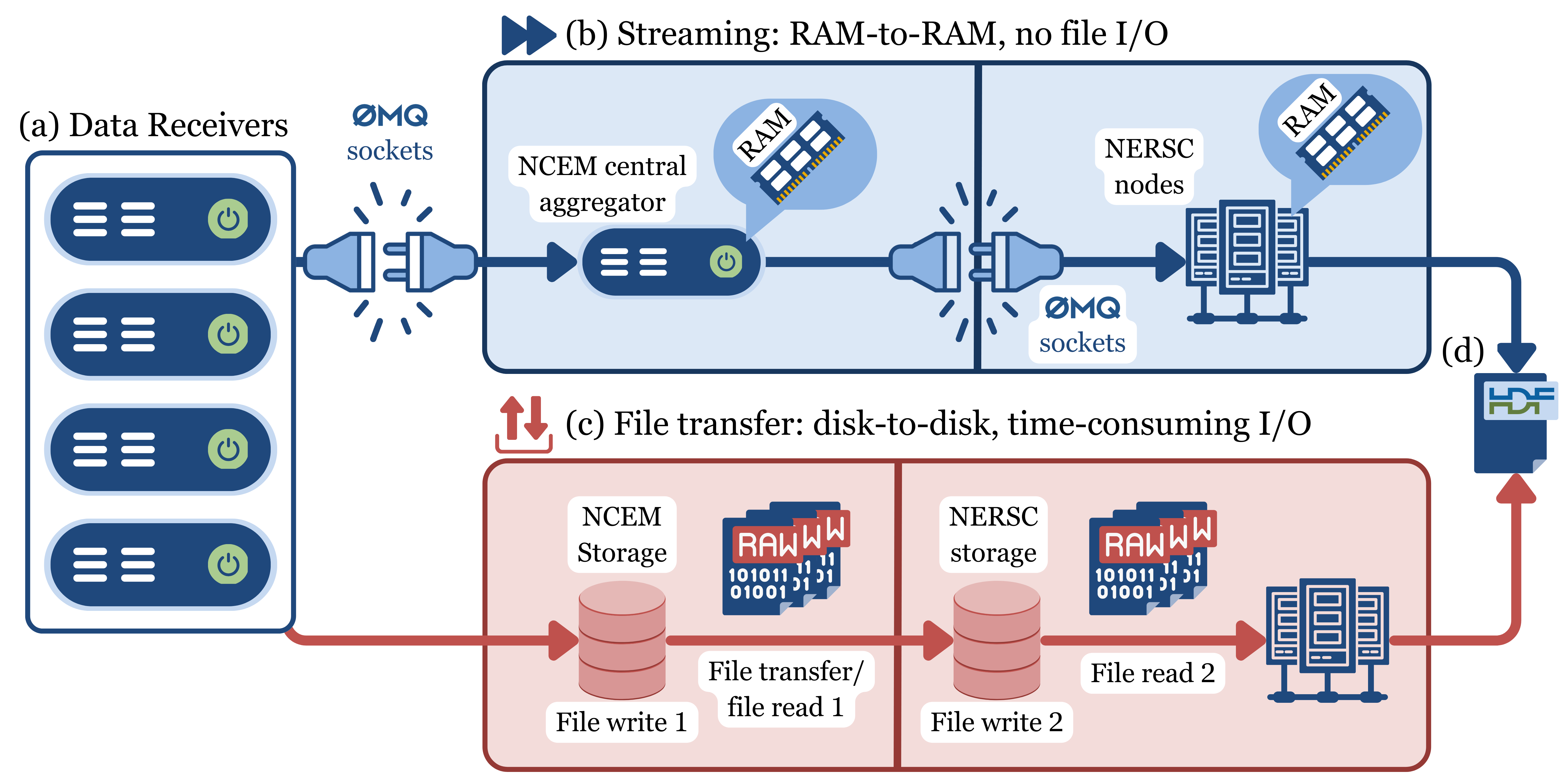}
    \caption{Schematic comparison of the data streaming pipeline (blue pathway, a-b-d) with the file transfer pipeline (red pathway, a-c-d).  Starting from the data receivers (a), the streaming approach employs \textit{ZeroMQ} sockets to bypass raw file disk storage at NCEM, enabling direct RAM-to-RAM transfer. Sockets are created on the data receivers, a centralized aggregator server at NCEM, and NERSC compute nodes to facilitate this transmission. Conversely, the file transfer approach requires several intermediate file storage operations to move the data from NCEM to NERSC. In both pathways, the thick vertical line indicates the network border between NCEM and NERSC. Using \textit{stempy}, the data is electron counted and saved in a single HDF5 file for further processing (d).}\label{fig:benefits_streaming} % chktex 13
\end{figure*}

The data receiving servers at NCEM (Fig.~\ref{fig:benefits_streaming}{a}) handle detector data retrieval, data formatting, and disk storage of raw data files (see background section). Traditionally, each server accumulates data for one sector of the detector in memory during a scan and writes it to disk as files (Fig.~\ref{fig:benefits_streaming}{c}) after acquisition is complete. We replaced this disk write operation with our \textit{ZeroMQ} streaming operation represented by the outlet socket attached to Fig.~\ref{fig:benefits_streaming}{a}. These sockets transmit the data from the server's RAM to a central aggregator server.

\subsection{Routing the Data to NERSC}
The aggregator server routes data to NERSC for frame reassembly and processing. Its sockets are graphically represented by the inlet socket attached to Fig.~\ref{fig:benefits_streaming}{b}. The routing strategy on the central aggregator uses sector metadata (the frame number) to forward data (outlet socket attached to the aggregator in Fig.~\ref{fig:benefits_streaming}{b}) to its corresponding node at NERSC (inlet socket attached to NERSC nodes in Fig.~\ref{fig:benefits_streaming}{b}). This data routing ensures equitable distribution of frames across the NERSC compute nodes, maintaining a consistent computational load across them. Further, it guarantees that all sectors of a given frame are routed to the same NERSC compute node --- sector data is initially dispersed among the receiving servers (see Fig.~\ref{fig:overview_prior_work}{b}), and they must be assembled on the same NERSC node before processing (see Fig.~\ref{fig:overview_prior_work}{g}).

\subsection{Live Electron Counting at NERSC}
At NERSC, the data is ingested into the compute nodes' RAM (Fig.~\ref{fig:benefits_streaming}{b}) from the upstream centralized aggregator. Full frames are automatically processed using the electron counting algorithm in the \textit{stempy} package~\citep{stempy}. After all frames have been received, the sparse, electron counted data is saved in a single HDF5 file (Fig.~\ref{fig:benefits_streaming}{d}). The entire system is now ready for another acquisition.

\section{Workflow from the User's Perspective}\label{sec:workflow}

Many users, particularly those without experience in HPC, may find the prospect of initiating a streaming job on a supercomputing cluster to be daunting. To address this, we extended the functionalities of \textit{Distiller}~\citep{distiller}. Prior to this work, \textit{Distiller} served as a web portal primarily for cataloging data sets, tracking metadata, and initiating processing jobs at NERSC. Our enhancements allow users to initiate a streaming compute job through the \textit{Distiller} web interface. Supplemental Video 2 demonstrates starting a session using the \textit{Distiller} interface and subsequently collecting several acquisitions~\citep{zenodo_videos}. % chktex 13

With a single mouse click in \textit{Distiller}, the necessary connections between NCEM and NERSC are automatically established. This enables users to focus on their experiments while the data is seamlessly streamed to NERSC. As a result, datasets are rapidly available for further analysis, eliminating user distraction and delays associated with manually starting a separate job for each dataset. % chktex 13

The user monitors the progress of their streaming session and initiates data analysis notebooks using NERSC's \textit{Jupyter} ecosystem directly from the \textit{Distiller} web interface (see Supplemental Video 3)~\citep{zenodo_videos}, which is enabled by NERSC's Superfacility API~\citep{thomas2017toward, thomas2021interactive, henderson2020accelerating,enders2020cross, parkinson2020interactive}. Integration of data acquisition, transfer, and analysis into a unified workflow enhances user productivity and enables more complex, data-intensive experiments. The streaming capability have been utilized on the TEAM 0.5 microscope for about 8 months providing streamlined data transfer and analysis for real user experiments. The code for \textit{Distiller} is publicly accessible and can be found in \cite{distiller}.

\section{Microscope Stability Experiment and Workflow Comparison}\label{sec:results}

\subsection{Stability Experiment}\label{sec:stability}

In order to explore the capabilities enabled by the streaming approach, we conducted a real experiment that mimics a typical high-throughput microscopy workflow --- the collection of data at regular time intervals for an extended period, hereafter referred to as a multi-scan experiment. The goals were three-fold: first, to generate a large volume of data that would challenge the streaming system's capabilities; second, to quantify the microscope's stability over time; and third, to show that the system can produce many high quality 4D-STEM datasets amenable to advanced analytical techniques, such as ptychography.

The experiment was performed on the aberration corrected TEAM 0.5 outfitted with the all piezo-electric TEAM Stage. This stage offers exceptional stability, with a nominal drift rate of 2 pm \(\mathrm{s}^{-1}\), and allows for tilting up to ±180\(^\circ\) within the 2.5 mm pole piece gap~\citep{ercius2012team}. The microscope was operated at an accelerating voltage of 300 keV, a convergence angle of 17.1 mrad, a sample tilt of 0\(^\circ\), a probe current of 20 pA, and a probe step size of 0.36 \AA. A standard sample made of gold nanoparticles with approximate diameter of 5-10 nm was prepared by chemical vapor deposition (CVD) of gold onto an ultra thin carbon substrate. Using an automated data collection script, we acquired 60 4D-STEM datasets at 55-second intervals (total duration of 55 minutes), each with dimensions of 512 \(\times\) 512 \(\times\) 576 \(\times\) 576. Each dataset consists of 173 GB of raw data, culminating in a total data volume of 10.4 TB streamed to NERSC. Each dataset was successfully acquired, transmitted to NERSC, reduced by electron counting, and stored for further analysis. The total data volume was reduced from 10.4 TB down to a more manageable size of 125 GB through counting. % chktex 21 chktex 13 chktex 8

\begin{figure*}[!t]
    \centering
    \includegraphics[width=1\linewidth]{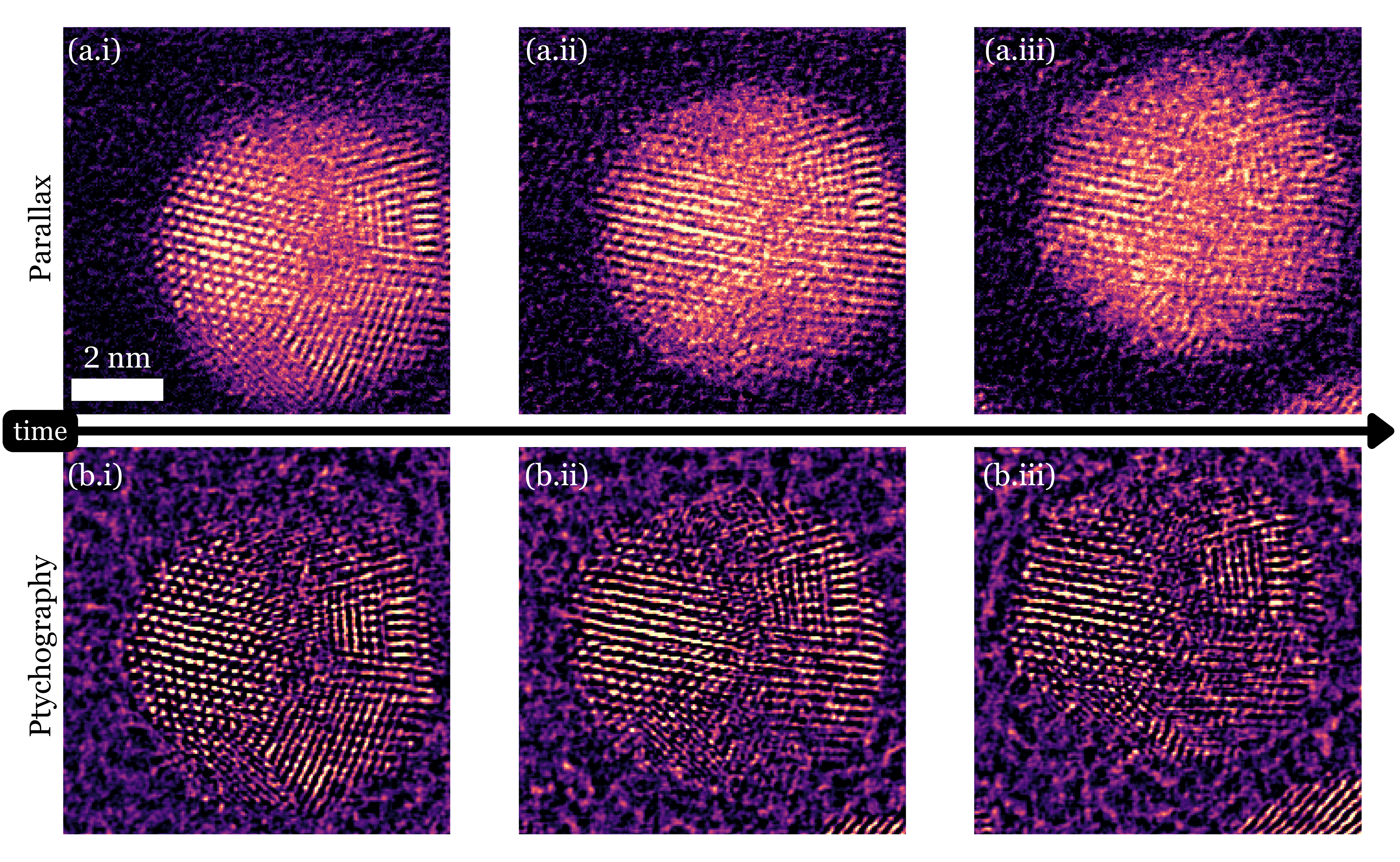}
    \caption{Reconstructions of the same gold nanoparticle using (a) parallax and (b) ptychography over the course of the nearly hour long experiment. (i), (ii), and (iii) display reconstructions from the experiment's start, middle, and end. % chktex 12
    }\label{fig:parallax_ptycho}
\end{figure*}

\begin{figure*}[!t]
    \centering
    \includegraphics[width=1\linewidth]{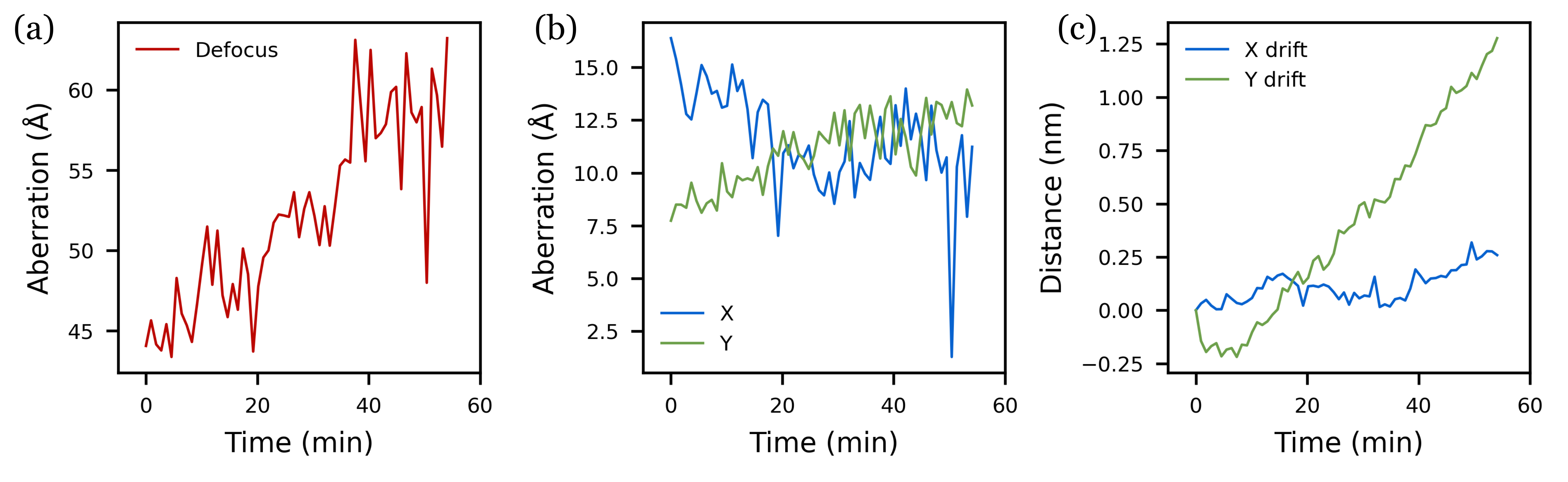}
    \caption{Fitted parameters from reconstructions in Fig.~\ref{fig:parallax_ptycho}. (a) Defocus (C1) drift of the TEAM 0.5 during the experiment, 0.3 pm \(\mathrm{s}^{-1}\). (b) Probe astigmatism (X and Y) drift throughout the experiment, both drifting at about 0.2 pm \(\mathrm{s}^{-1}\). (a) and (b) were both fit using the data in Fig.~\ref{fig:parallax_ptycho}{a}. (c) Lateral drift of the sample, fit with cross-correlation using the first ptychography reconstruction as the basis \( \left( x = 0, \, y = 0 \right) \). % chktex 12
    }\label{fig:aberrations} 
\end{figure*}

The large number of high quality 4D-STEM scans acquired during this hour long experiment provides an opportunity to measure changes in the microscope using advanced techniques such as parallax or tilt-corrected bright field and ptychography~\citep{varnavides2023iterative, yu2024dose}. Here, we used the \texttt{ParallaxReconstruction} and \texttt{SingleslicePtychographicReconstruction} classes in py4DSTEM version 0.14.3~\citep{savitzky2021py4dstem, varnavides2023iterative} to perform the reconstructions on each dataset in the time series (see our accompanying data repository in~\cite{this_paper_code_repo} for the reconstruction settings along with an example \textit{Jupyter} notebook for one of the acquisitions). Representative (a) parallax and (b) ptychography reconstructions at the beginning (i), middle (ii), and end (iii) of the series are shown in Fig.~\ref{fig:parallax_ptycho}. These reconstructions indicate that atomic resolution is maintained throughout the experiment, owing in part to the exceptional stability of the TEAM 0.5 microscope and stage --- we made no adjustments to the microscope during the experiment. 
 
We quantify the microscope's stability by inspecting the estimated microscope parameters from the parallax and ptychography results. In a parallax reconstruction, virtual images from different positions in the bright field disk are aligned through cross-correlation. The image shifts are imparted on these virtual images based on the gradient of the aberration surface of the incoming beam and the rotation between real and reciprocal space in the microscope setup. By fitting the aberration profile of these shifts and rotations, we can estimate changes in low order aberrations during the course of the experiments.  Fig.~\ref{fig:aberrations}{a} shows the estimated change in defocus over the full hour of data acquisition, amounting to a drift rate of 0.5 pm \(\mathrm{s}^{-1}\), which is either due to stage or lens drift. The defocus drift value is not typically measured due to the projection nature of the STEM. % chktex 13

We also expect other aberrations to change during the course of the experiment due to lens drift~\citep{intrinsic_instability_ACEM}, and we can estimate the probe astigmatism (A1) in X and Y for all 60 datasets (Fig.~\ref{fig:aberrations}{b}) based on the A1\(_X\) and A1\(_Y\) determined from the probe estimate over time. Both astigmatism directions have a drift rate of approximately 0.2 pm \(\mathrm{s}^{-1}\).  Iterative electron ptychography can be used to solve for the object as well as the probe from a 4D-STEM dataset and produces a high-resolution and high signal-to-noise reconstruction~\citep{varnavides2023iterative}. We further quantified the lateral drift of the sample by employing cross-correlation techniques on the high-resolution ptychographic reconstructions. The sample exhibited a movement of approximately 1.3 nm in the positive Y direction and around 0.3 nm in the positive X direction (Fig.~\ref{fig:aberrations}{c}). These shifts are well within the published stability limits of the microscope stage, which allows for a maximum drift of 6.6 nm over the 55-minute experiment duration, as calculated from the stage's drift rate of 2 pm \(\mathrm{s}^{-1}\). 

\subsection{Workflow Comparison}\label{sec:workflow-comparison}

\begin{figure*}[!t]
  \centering
  \includegraphics[width=1\linewidth]{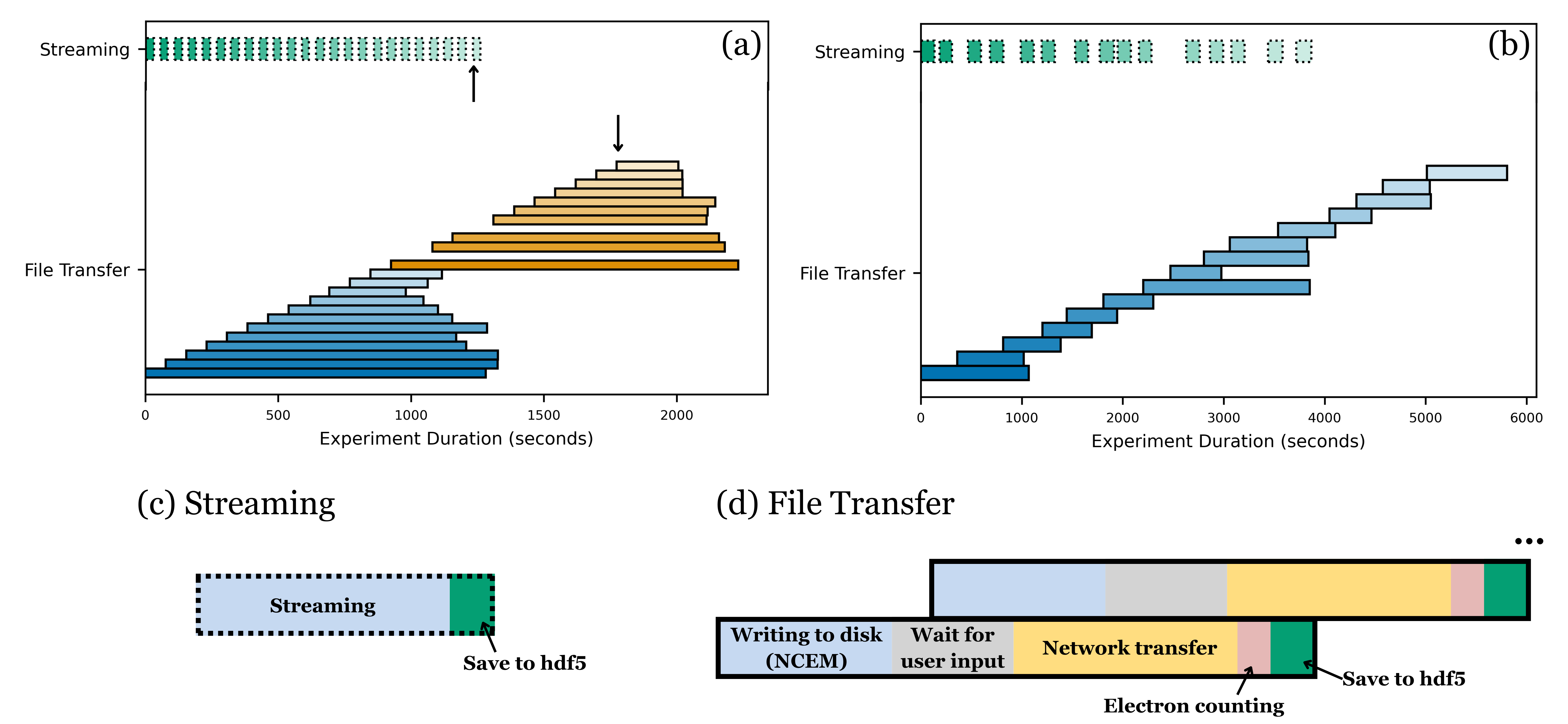}
  \caption{Timeline diagrams of the streaming workflow compared to the file transfer workflow for two different types of experiments: (a) multi-scan, where data is automatically acquired at regular intervals similar to the stability experiment; and (b) 4D-STEM tomography, where data is collected at semi-regular intervals, but adjustments must be made to the microscope between acquisitions. The left side of each horizontal bar represents the acquisition start time, and the right side indicates the time when the electron-counted data is available at NERSC. (c) and (d) qualitatively indicate the serial steps taken within each of these bars for streaming and file transfer, respectively. The arrows in (a) represent the twenty-fourth acquisition in both workflows.}\label{fig:timeline} % chktex 13
\end{figure*}

In our recent work~\citep{welborn2024accelerating}, we established that streaming a dataset from NCEM to NERSC is five- to fourteen-fold faster than the conventional file transfer workflow in terms of raw throughput of raw detector data without counting electron events (i.e., the electron beam was off). Here, we will compare these two workflows through an analysis of historical data from four real user experiments with electron events, as illustrated by the timelines in Fig.~\ref{fig:timeline}. File transfer experiments exhibit extended durations due to concurrent dataset transmissions and manual user interactions with \textit{Distiller}, which introduce delays. Conversely, streaming maintains consistent and reliable transfer times, making data immediately accessible at NERSC post-acquisition. 

To construct these plots, we determined the last-modified timestamps for two key files created on the NERSC file system for each acquisition. The `start time' is marked by the timestamp of the simultaneously acquired HAADF-STEM image file uploaded to NERSC immediately at the end of each acquisition, and the `end time' by the timestamp of the electron-counted data file. This pair forms one of the horizontal bars in Fig.~\ref{fig:timeline}. It is important to note that these timestamps are synchronized using the same clock, avoiding timing discrepancies across different devices in the distributed workflow environment. For clarity, we do not detail the intermediate steps between the start and finish in Fig.~\ref{fig:timeline}{a-b}, instead displaying representative timeline snippets in Fig.~\ref{fig:timeline}{c-d}. For the file transfer workflow (Fig.~\ref{fig:timeline}{d}), there are five serial steps: writing the data to the local drive at NCEM (light blue), waiting for user input to initiate network transfer (grey), the network transfer to NERSC file system (yellow), followed by electron counting (pink), and finally saving to hdf5 (green). Conversely, there are only two steps in the streaming workflow: streaming (light blue), followed by saving to hdf5 (green). 

In Fig.~\ref{fig:timeline}{a}, we compare the file transfer and streaming workflows for multi-scan experiments similar to the stability experiment described above. Each acquisition amounted to 173 GB of raw data, with data dimensions of 512 \(\times\) 512 \(\times\) 576 \(\times\) 576. During the experiment using the file transfer workflow, the user allowed a batch of acquisitions to accumulate on the NCEM file system and then initiated many NERSC transfer jobs using \textit{Distiller} in reverse-acquisition order. This results in a pyramid-shaped timeline for each batch, since the most recent acquisition was transferred to NERSC first. In Fig.~\ref{fig:timeline}{a}, two batches are shown: the first starting with the dark blue bar and ending with the light blue bar, and the second starting with the dark orange bar and ending with the light orange bar. Notably, the second batch exhibits gaps indicating missing acquisitions. These omissions could either be deliberate, perhaps due to adjustments in the microscope setup causing concerns with these acquisitions, or unintentional due to transfer failures. In either case, this underscores the disadvantages of having a human in the loop for repetitive file transfer tasks, as real-time decision-making distracts from the ongoing experiment. % chktex 21

The delay between acquisition and data availability is significantly longer in the file transfer workflow compared to streaming. For instance, in the first batch, the user waited 20 minutes for the initial acquisition to be available (dark blue bar in Fig.~\ref{fig:timeline}{a}). Even the last acquisition in the first batch, one of the shortest timeline bars in the series, required about 270 seconds to become accessible---almost an order of magnitude slower than the consistent 30-second time to processed data observed in the streaming workflow. Our previous work showed that the average file transfer duration for similar sized data sets is approximately 139 seconds (refer to the Results section of reference \citenum{welborn2024accelerating}). However, that analysis did not account for additional overhead found in real experiments, such as simultaneous data transfer and acquisition, Perlmutter queue times, and user interactions needed to initiate transfers in \textit{Distiller}. Together, these delays amounted to doubling the waiting period for the microscope user. Conversely, the streaming acquisitions, represented in teal, were available approximately 30 seconds after each acquisition as no human interaction is required between acquisitions, and simultaneous data transfer and acquisition do not occur. The arrows in Fig.~\ref{fig:timeline}{a} point to the last (twenty-fourth) acquisition in both series, including the omitted file transfer acquisitions, indicating the streaming workflow enables collection of data at a faster rate. % chktex 21

In Fig.~\ref{fig:timeline}{b}, we compare the workflows for a 4D-STEM tomography experiment, where a user spends time between acquisitions to align the sample and microscope at a set of rotation angles. Each acquisition amounted to 695 GB of raw data, with data dimensions of 1024 \(\times\) 1024 \(\times\) 576 \(\times\) 576. Here, the user was able to tilt, center, and focus the object in the field of view faster than the file transfer pipeline was able to produce processed data. The user thus had to wait for the NERSC process to complete before acquiring a new scan. Further, the user was required to initiate file transfers (disrupting their focus on the experiment) and monitor the file transfer process during the experiment to avoid overtaxing the system. Conversely, the streaming workflow (initiated with one interaction at the start of the experiment) produced finalized data before the next scan was initiated indicating processing time was less than microscope operation time. % chktex 21

In both cases, it is clear that more data can be acquired in a shorter amount of time with better consistency. There is also additional benefit in removing several extra steps from the experimental workflow, especially the need for users to initiate processing jobs.

\section{Conclusions}\label{sec:conclusions}

In this work, we demonstrate the advantages of a streaming-based data transfer workflow over traditional file-based workflows, which often suffer from performance bottlenecks due to disk I/O operations. By bypassing local and remote disk I/O and transferring data directly over the network to a HPC center, our pipeline enables on-the-fly processing on remote hardware with better capabilities. This streaming pipeline seamlessly connects a high frame rate direct electron detector (the 4D Camera) to an HPC center (NERSC). 

The pipeline's capabilities were demonstrated through an hour-long experiment, where 60 4D-STEM datasets totalling over 10 TB of raw data were acquired, streamed, and electron-counted in real time at NERSC, resulting in a compressed data size of 125 GB. This experiment not only tested the streaming workflow's ability to handle large volumes of data but also evaluated the entire system's capacity to produce large numbers of high-quality datasets suitable for advanced analyses such as parallax and ptychography. % chktex 13

A key benefit of our streaming approach, beyond the already-established increase in raw throughput~\citep{welborn2024accelerating}, is the significant reduction in human interaction required. Our comparative analysis of historical data from real user experiments reveals that automating the data transfer process increases throughput, minimizes the potential for human error, and eliminates the overhead associated with manual interactions.

Furthermore, the user focused design of our solution abstracts away the complexities of HPC, allowing researchers to focus on scientific inquiry rather than the intricacies of computation. This streaming system is integrated into the \textit{Distiller} web frontend, simplifying the workflow. The system is in daily use on the TEAM 0.5 microscope at NCEM. % chktex 13

This work represents an important step forward in the integration of HPC resources with EOS facilities, addressing critical challenges in data management and computational efficiency. It serves as a model for similar integrations in other data-intensive scientific domains, having implications that extend beyond the immediate context of one electron microscopy detector. Future work will focus on expanding its applicability to other experimental setups and analytical techniques.

%%%%%%%%%%%%%%

\section{Competing interests}
No competing interest is declared.

\section{Acknowledgments}
Work at the Molecular Foundry was supported by the Office of Science, Office of Basic Energy Sciences, of the U.S. Department of Energy under Contract No. DE-AC02-05CH11231. This research used resources of the National Energy Research Scientific Computing Center (NERSC), a U.S. Department of Energy Office of Science User Facility located at Lawrence Berkeley National Laboratory, operated under Contract No. DE-AC02-05CH11231 using NERSC awards BES-ERCAP0024753 and BES-ERCAP0024754. We would like to thank Gatan, Inc.\ as well as P Denes, A Minor, J Ciston, J Joseph, Vamsi Vytla, and I Johnson who contributed to the development of the 4D Camera. We would also like to thank A Saha and A Bhalla-Levine for compiling experimental information for use in Figure 5. %chktex 8

\end{document}